\documentstyle[psfig,12pt]{article}
\def\1{\hbox{{1}\kern-.25em\hbox{l}}}
\begin{document}
\begin{center}{\Large
Does one observe chiral symmetry restoration in baryon spectrum?}
\end{center}
\bigskip
\begin{center}{\large Thomas D. Cohen$^a$ and Leonid Ya. Glozman$^b$}
\end{center}
\medskip
{ \it $^a$ Department of Physics, University of Maryland,
College Park, Maryland 20742-4111,USA}\\
{ \it $^b$Institute for Theoretical
Physics, University of Graz, Universit\"atsplatz 5, A-8010
Graz, Austria}

\begin{abstract}
It has recently been suggested that the parity doublet
structure seen in the spectrum of highly excited baryons
may be due to effective chiral symmetry restoration for
these states. We review the recent developments in this
field. We demonstrate with a simple quantum-mechanical
example that it is a very natural property of quantum systems
that a symmetry breaking effect which is important for
the low-lying spectrum of the system, can become unimportant
for the highly-lying states; the highly lying states reveal
a multiplet structure of nearly degenerate states. Using the
well established concepts of quark-hadron duality, asymptotic
freedom in QCD and validity of the operator product expansion
in QCD we show that the spectral densities obtained with the
local currents that are connected to each other via chiral
transformations, very high in the spectrum must coincide.
Hence effects of  spontaneous breaking of chiral symmetry
in QCD vacuum that are crucially important for the low-lying spectra,
become irrelevant for the highly-lying states. Then to the extent that
identifiable hadronic resonances still exist in the continuum
spectrum at high excitations this implies that the highly
excited hadrons must fall into multiplets associated with the
representations of the chiral group.  We demonstrate that this
is indeed the case for meson spectra
in the large $N_c$ limit. All possible parity-chiral multiplets are
classified for baryons and it is demonstrated that the existing
data on highly excited $N$ and $\Delta$ states at masses of
2 GeV and higher is consistent with approximate chiral symmetry
restoration. However  new experimental studies are needed to
achieve any definitive conclusions.
\end{abstract}

\bigskip
\bigskip
\bigskip
\bigskip
\noindent
------------------------------------------------------------------------------

\noindent
Review commissioned by {\bf Int. J. of Modern Physics A}
\newpage

\section{Introduction}

QCD is a very strange theory. On the one
hand, it is formulated in terms of quarks and
gluons; on the other hand, these particles
are never observed experimentally. Due to confinement,
a fundamental but poorly understood property of the
theory, only color neutral
(color singlet) particles are possible as asymptotic states.
Thus only hadrons will hit our detectors but not quarks or gluons.\\

Another interesting property of QCD is that its Lagrangian
has an
almost perfect $SU(2)_L \times SU(2)_R$ chiral symmetry,
which is broken only by the very small up and down quark masses.
However this symmetry is not directly
observed in the world - it is hidden, {\it i.e.} spontaneously broken.\\

One of the most intriguing fields of study
in physics, both experimentally and theoretically,
is the exploration of regimes where the
color degree of freedom is not trapped into
a small volume of hadrons, and where
chiral symmetry is restored. These regimes
are expected to be achieved as phases of bulk matter
at high temperature
or/and at high density.  Experimentally it is
believed these conditions can be reached in heavy ion
collisions.\\

As has been argued very recently \cite{G1,TG}, however, it is possible that
a regime exists where the chiral symmetry is (almost) restored
 but  hadrons as entities
still exist. To see this regime one needs to only 
study very highly excited hadrons.  Such a task
is experimentally feasible with  present facilities and
with a careful analysis of older data.  Evidence of effective
chiral restoration might then be seen in the spectroscopic
patterns of the  highly excited hadrons.
It is amusing to note, that  data which hint
the onset of this regime have existed for many years, but
have not attracted much attention.\\

The aim of this paper is to review the recent developments
in this field. As will be seen very little can be
calculated directly.   However we can reach some
definitive conclusions on the
asymptotic symmetry properties of spectral functions using
only very
general theoretical grounds such as
the well-established idea of quark-hadron
duality, the property of asymptotic freedom and
the apparatus of operator product expansion (OPE)
in QCD. In particular
one can show that the effects of spontaneous symmetry breaking
must smoothly switch off once we go up in the spectrum.
This then implies that the spectrum of highly excited states
should reveal the chiral symmetry of QCD.  This will be
reflected in multiplet structures for the highly excited
states. There is a caveat which must be made at this point.
On theoretical grounds we have no {\it a priori} way to establish
whether or not hadron states remain as identifiable resonances
when one studies the spectrum high enough for chiral symmetry
to be effectively restored.  We conjecture, however, that it is
possible, and then ask whether the experimental data support this
conjecture.  Moreover we have strong theoretical
evidence that the conjecture is not impossible:
in large $N_c$ QCD the meson spectrum can be shown to
behave in precisely this way.\\

The idea that the fundamental strong interaction theory
should posses an approximate $SU(2)_L \times SU(2)_R$
(or $SU(3)_L \times SU(3)_R$) chiral symmetry dates back
to the 1960s  \cite{GML,ADDA,P}. One of the most important insights
from this was that this symmetry must be spontaneously broken
in the vacuum (i.e. realized in the Nambu-Goldstone mode).
The most important early arguments were: (i) the absence of
parity doublets in the hadron spectrum (if the chiral symmetry
were realized in the Wigner-Weil mode---{\it i.e.} if the vacuum were
trivial---then the hadron spectrum would have to reveal the multiplets
of the chiral group which are manifest as parity doublets);
 (ii) the exceptionally low mass of pions,
which are taken to be pseudo-Goldstone bosons associated with
the spontaneously broken axial symmetry.
Substantial phenomenological work in this field occurred during the
1960s.  However, the microscopic foundations of the symmetry
were not well understood.
When QCD appeared in the 1970s, one of the reasons for its rapid
acceptance was that it very naturally explained all of the
successes  of chiral current algebra.\\

In parallel with the development of the
current algebra, the physics of  excited hadron states
also attracted significant attention. One important
result for the present context is that the
famous linear-like behavior of Regge trajectories
required the parity doubling of baryonic states, due
to the so called generalized McDowell symmetry \cite{Collins}.
And indeed, the ``Regge physicists'' have observed that
high in the baryon spectrum there appear more and more
parity doublets. On the other hand, such doublets were
clearly absent low in the spectrum, a fact which could
not be understood from the Regge physics perspective.\\

Given these facts, it is difficult to understand
why the simple idea that chiral symmetry is effectively
restored for states high in the spectrum was not been explored
long ago. From the present perspective there is no
contradiction between the linear-like behavior
of the highly-lying baryon Regge trajectories and the
absence of parity doublets low in the spectrum. We know,
that the well established spontaneous breaking of chiral
symmetry prevents the low-lying states from doubling.
This, together with the McDowell symmetry suggests then
that there should be no systematic linear parallel
Regge trajectories for positive and negative parity states
low in the baryon spectrum.  Of course, this is precisely what is seen
\cite{GREGGE}.\\

To the best of our knowledge the first speculations that the parity
 doublets seen in the highly
lying baryon states probably reflect the chiral symmetry restoration
have appeared only recently \cite{GR}. This possibility was
taken seriously in ref. \cite{G1}. It was argued 
in the latter work that it is
quite natural to expect chiral symmetry restoration high
in the spectrum because in this case the typical momenta
of quarks should be high, and once it is high enough and
approach the chiral symmetry restoration scale the
dynamical (constituent) mass of quarks should drop off and
as a consequence the chiral symmetry should be restored.
In other words, at high momenta the valence quarks in
hadrons should decouple from the quark condensates. This
smooth chiral symmetry restoration is seen by the
presence of approximate
parity doublets high in the spectrum.
This perspective, while interesting in its own right,
suggests strong limitations on constituent quark models.
  Thus it
 becomes
evident that the constituent quark model is not applicable
high in the spectrum.\\

That chiral symmetry must indeed be restored high
in the spectrum was shown in ref. \cite{TG}. This
can be seen directly from quark-hadron duality, asymptotic freedom property
of QCD and the OPE. It was shown that even if the chiral
symmetry is strongly broken in the vacuum, and hence
 the low-lying states do not manifest chiral
symmetry, one should expect effective chiral symmetry
restoration in the spectral density for highly lying states in the spectrum.
This then suggests that the highly excited hadrons should
fall into multiplets of nearly degenerate states
which are associated with representations of the chiral group.
All such possible
multiplets have been classified and it was demonstrated
that the existing data on highly lying baryon resonances
are appear to be compatible with this. New experimental studies
are needed, however, in order to make any definite statements
whether we see approximate chiral symmetry restoration at
baryon masses of 2 GeV and higher. If it does take a place,
then the spectrum of highly excited baryons should consist
exclusively of approximate parity doublets.\\

These ideas have immediately been extended by Beane \cite{Beane}
for highly excited vector and axial vector mesons, which also
should be degenerate once the chiral symmetry is restored.
Beane has proved that it is impossible in the large $N_c$
limit to satisfy the chiral symmetry of QCD Lagrangian
and the quark-hadron duality if the highly excited vector
and axial vector mesons do not form parity doublets.
Yet, there appear to be no experimentally observable parity doublets
in this meson case, in contrast to  baryons. \\

We should stress that this smooth chiral
symmetry restoration should not be confused with a phase
transition. If one defines a phase transition in the usual way
as an abrupt
transition of the vacuum from one phase to the other, then
the phase transition
implies symmetry restoration through the whole spectrum of
the system, not only for the highly lying states.
On the contrary, the property
which we discuss,  is a particular case of a different, but
quite general, physical property. Namely, if one studies
 a system (in our case the QCD vacuum)
with
a  high frequency (or short distance) probe that is sensitive to
distances that are much smaller than the  length
associated with the symmetry breaking in the system,
then the response of the system to this probe is  essentially 
the same as if
there were no symmetry breaking in the system. For example,
if one probes a metal in the superconducting phase with photons
$\hbar \omega \gg  \Delta$, then the superconducting coherence
in the ground state (which is analogous with the QCD vacuum
for our case) become unimportant and response of the
superconductor is the same as of normal metal. There is a smooth transition
from the regime $\hbar \omega \sim 2\Delta$---where the effects of
the superconducting (phase coherent) structure of the metal
 are crucially important, to the
regime $\hbar \omega \gg  \Delta$---where they are not.\\

This short review consists of a number of sections. In the second
one we give a short overview of chiral symmetry in QCD, which may
be omitted by any reader who is familiar with the subject.
In the third section we give a simple pedagogical quantum mechanical
example. We show that it is a very natural property of
quantum systems that a symmetry breaking effect which is important
for the low-lying spectrum of the system, can become unimportant
for the highly-lying states; the high lying  states reveal
 a multiplet structure of nearly degenerate states.
In the next section we discuss
a tool, the spectral density, which can be used to study systems 
with continuous
spectra.  In
that section, we show why one should expect the smooth chiral symmetry
restoration high in the spectrum. The fifth section is devoted
to somewhat delicate question about to what extent one
can use the language of highly excited resonance states once
we are in the continuum. In the sixth section we classify
all possible parity-chiral multiplets that should be recovered
once we approach the regime of chiral symmetry restoration.
The experimental data for both low-lying baryon states and
the highly lying states are analyzed in the seventh section,
where we show that the existing experimental pattern of $N$
and $\Delta$ excitations is such that it can be interpreted
that one achieves the regime of approximate chiral symmetry
restoration at baryon masses of 2 GeV. Nevertheless, we stress
that the new experimental studies are called for to make
this conjecture a more definite statement. Finally we present
a general outlook in a concluding section.\\

\section{Short overview of chiral symmetry in QCD}

In the chiral limit the quarks are massless. In reality
the masses of $u$ and $d$ quarks are quite small  compared to
the typical hadronic scale of 1 GeV; to a good approximation
they can be neglected. In this limit the right and left components
of quark fields

\begin{equation}
\psi_R = \frac{1}{2}\left( 1+\gamma_5 \right ) \psi,~~
\psi_L = \frac{1}{2}\left( 1-\gamma_5 \right ) \psi
\label{RL}
\end{equation}

\noindent
are decoupled. This is because the quark-gluon interaction
is vectorial, 
$\bar \psi \gamma^\mu \psi A_\mu$, which does not mix the right- and
left-handed components of quark fields. On the other hand in
the chiral limit the quark-gluon interaction is insensitive
to the specific flavor of quarks.  For example one can substitute,
 the $u$ and $d$ quarks by  properly normalized orthogonal 
linear combinations of $u$ and $d$
quarks ({\it i.e.} one can perform a rotation in the flavor space)
and nothing will change. Since the left- and right-handed components
are completely decoupled, one can perform two independent flavor
rotations of the left- and right-handed components:

\begin{equation}
\psi_R \rightarrow 
\exp \left( \imath \frac{\theta^a_R \tau^a}{2}\right)\psi_R; ~~
\psi_L \rightarrow 
\exp \left( \imath \frac{\theta^a_L\tau^a}{2}\right)\psi_L,
\label{ROT}
\end{equation}

\noindent
where $\tau^a$ are the isospin Pauli matrices and the
angles $\theta^a_R$ and $\theta^a_L$ parameterize rotations
of the right- and left-handed components. These rotations leave
the QCD Lagrangian invariant. The symmetry group of these
transformations,

\begin{equation}
SU(2)_L \times SU(2)_R,
\label{chsymm}
\end{equation}

\noindent
is called chiral symmetry.\\

Now generally if the Hamiltonian of a system is invariant under
some transformation group $G$, then one can expect
that one can find states which are simultaneously eigenstates
of the Hamiltonian and of the Casimir operators of the group , $C_i$.
Now, if the ground
state of the theory, the vacuum, is invariant under the same
group, {\it i.e.}  if for all $U \in G$
\begin{equation}
 U | 0 \rangle = | 0 \rangle ,
\label{vac}
\label{symvaccond}\end{equation}
then eigenstates of this Hamiltonian corresponding to excitations
above the vacuum can be grouped into degenerate multiplets corresponding
to the particular representations of $G$. 
 This mode of symmetry is
usually referred to as the Wigner-Weyl mode.  Conversely, if 
(\ref{symvaccond}) does not hold,  the excitations do not
generally form degenegerate multiplets in this case.  This situation
is called spontaneous symmetry breaking.\\ 

If chiral symmetry were realized in the Wigner-Weyl mode, then the excitations
would be grouped into representations of the chiral group. 
The representations of the chiral group are discussed in
detail in  one of the following sections. The important
feature is that the every representation except the trivial one 
(which by quantum numbers cannot include baryons) necessarily implies
parity doubling. In other words, for every baryon with
the given quantum numbers and parity, there must exist another
baryon with the same quantum numbers but opposite parity and
which must have the same mass.
In the case of mesons the chiral representations combine, e.g.
the vector mesons with the axial vector mesons, which should
be degenerate. This feature is definitely not observed
for the low-lying states in hadron spectra. This means that
eq.~(\ref{symvaccond}) does not apply;
the chiral symmetry of QCD Lagrangian is spontaneously (dynamically)
broken in the vacuum, {\it{i.e.}} it is hidden. Such a mode
of symmetry realization is referred to as the Nambu-Goldstone one.
\\

The independent left and right rotations (\ref{ROT}) can be
represented equivalently  with independent isospin and axial 
rotations

\begin{equation}
\psi \rightarrow 
\exp \left( \imath \frac{\theta^a_V \tau^a}{2}\right)\psi; ~~
\psi \rightarrow 
\exp \left( \imath  \gamma_5 \frac{\theta^a_A\tau^a}{2}\right)\psi.
\label{VA}
\end{equation}

In the Wigner-Weyl mode, the invariance under
these transformations implies three conserved vector and three conserved
axial vector currents. The existence of 
 approximately degenerate isospin multiplets
in hadron spectra suggests that the vacuum is invariant under
the isospin transformation. Indeed, from the theoretical side the Vafa-Witten
theorem \cite{VW} guarantees that  in the local gauge theories
the vector part of chiral symmetry cannot be spontaneously broken.
The axial  transformation mixes
states with opposite parity. The fact that all states
do not have parity doublets
implies that the vacuum is not invariant under the
axial  transformations. In other words the almost perfect
chiral symmetry of the QCD Lagrangian is dynamically broken
down by the vacuum to the vectorial (isospin) subgroup

\begin{equation}
SU(2)_L \times SU(2)_R \rightarrow SU(2)_I.
\label{breaking}
\end{equation}

The noninvariance of the vacuum with respect to the three axial
 transformations requires existence of three massless
Goldstone bosons, which should be pseudoscalars and form an
isospin triplet. These are identified with pions. The nonzero
mass of pions is entirely due to the {\it explicit} chiral symmetry breaking
by the small masses of $u$ and $d$ quarks. These small masses can
be accounted for as a perturbation. As a result the squares of
the pion masses are proportional to the  $u$ and $d$ quark masses
\cite{GOR}

\begin{equation}
m_{\pi^{+,-}}^2 = -\frac{1}{f_{\pi}^2} \frac{m_u + m_d}{2} 
(\langle \bar u u \rangle + \langle \bar d d \rangle) + O (m_{u,d}^2),
\label{gor1}
\end{equation}

\begin{equation}
m_{\pi^{0}}^2 = -\frac{1}{f_{\pi}^2} \left (m_u \langle \bar u u \rangle
 + m_d \langle \bar d d \rangle \right) + O (m_{u,d}^2).
\label{gor2}
\end{equation}

That the vacuum is not invariant under the axial transformation
is directly seen from the nonzero values of the quark condensates,
which are an order parameter for spontaneous chiral
symmetry breaking. These condensates are the vacuum expectation
values of the $\bar \psi \psi = \bar \psi_L \psi_R + \bar \psi_R \psi_L$
operator and at the renormalization scale of 1 GeV they approximately are

\begin{equation}
\langle \bar u u \rangle  \simeq \langle \bar d d \rangle  \simeq -
 (240 \pm 10 MeV)^3.
\label{con}
\end{equation}

\noindent
The values above are deduced from phenomenological considerations \cite{SR}.
Lattice
gauge calculations also confirm the nonzero and rather large values
for quark condensates. However, the quark condensates above are not
the only order parameters for chiral symmetry breaking. There exist
chiral condensates of higher dimension (vacuum expectation values of
more complicated combinations of  $\bar \psi$  and $ \psi$ that are
not invariant under the axial transformations). Their numerical values
are difficult to extract from phenomenological data, however, and
they are still unknown.\\

To summarize this section. There exists overwhelming evidence that
the nearly perfect chiral symmetry of the QCD Lagrangian is
spontaneously broken in the QCD vacuum. Physically this is because
 the vacuum state in QCD is highly nontrivial which can be seen
by the condensation in the vacuum state of the chiral pairs. These condensates
break the symmetry of the vacuum with respect to the axial transformations
and as a consequence, there is no parity doubling in the low-lying spectrum.
However, as we shall show, the role of the
chiral symmetry breaking quark condensates becomes progressively 
less important 
once we go up in the spectrum, {\it i.e.} the chiral symmetry is
effectively restored, which should be evidenced by the systematical appearance
of the approximate parity doublets in the highly lying spectrum. This is the
subject of the following sections.

\section{A simple pedagogical example}

One key theoretical idea underlying this review is that effect of spontaneous
chiral symmetry breaking on the spectrum becomes progressive less important as
one studies states higher in the spectrum.  As will be discussed later,
 this can
be seen directly from QCD.  It is none-the-less
instructive to consider a simple quantum mechanical
 system with some underlying symmetry that is
broken and accordingly is not readily apparent in the low lying spectrum but
which is effectively restored to good approximation for high lying states.
In
this section we will discuss such a system in the context of single particle
quantum mechanics.  This example will illustrate how this general phenomenon
can
come about.  The example we consider is a two-dimensional harmonic oscillator
(with an underlying $U(2)$ symmetry) with an added strong symmetry breaking
term. We choose the harmonic oscillator only for simplicity; the property
that will be discussed below is quite general one and can be seen in systems
with other type of symmetry.\\

Before discussing the example in any detail we wish to stress
that the example
is not in perfect analogy to the problem of interest in a number
 of  ways.
In the first place, the symmetry breaking in our pedagogical
example is explicit
while in the case of chiral symmetry breaking in QCD the principal
 effect of symmetry breaking on the spectrum is due to spontaneous
  symmetry breaking. Secondly, in our pedagogical example the
  spectrum is discrete while in the QCD case the spectrum is
  continuous.  Thus, one of the essential question in the QCD
  case---``Are the states still resonant when they are high
  enough in the spectrum for symmetry breaking to effective
  turn off?''---simply does not arise in this section. However, the example
  will make clear one essential thing: it is quite possible to
  have a system in which a symmetry breaking effect destroys the
  effect of a symmetry for the low lying spectrum (as seen by the
  lack of a multiplet structure) while to very good approximation
  the symmetry is manifest high in the spectrum.\\

The unperturbed system we consider is a two dimensional harmonic
oscillator.  We can always choose our units of time such that the
 vibration frequency is unity and our units on distance so that
 the spring constant is unity.  The Hamiltonian for such a system is

\begin{equation}
H_{\rm HO} \, = \, \frac{1}{2}
\left ( p_x^2 + p_y^2 + x^2 + y^2 \right ) \; \; .
\label{HO}
\end{equation}

\noindent
This Hamiltonian can be rewritten in terms of creation and
annihilation operators

\begin{equation}
a_x=\frac{1}{\sqrt {2}}(x+ \imath p_x), ~~
a_x^+=\frac{1}{\sqrt {2}}(x- \imath p_x);
\label{ax}
\end{equation}

\begin{equation}
a_y=\frac{1}{\sqrt {2}}(y+ \imath p_y), ~~
a_y^+=\frac{1}{\sqrt {2}}(y- \imath p_y)
\label{ay}
\end{equation}

\noindent
with
\begin{equation}
H_{\rm HO} \, = \, a_x^+a_x + a_y^+ a_y + 1.
\label{sq}
\end{equation}

The Hamiltonian above is a quadratic form in both $x$ and $p$
and as such  is invariant under  $U(2)$
(or equivalently $SU(2)\times U(1)$) transformations:

\begin{equation}
\left  ( \begin{array}{c}
x \, + \, i p_x \\
y \, + \, i p_y \end{array} \right ) \rightarrow
\left  ( \begin{array}{c}
x' \, + \, i {p'}_x \\
y' \, + \, i {p'}_y \end{array} \right ) \, =
U \, \left  ( \begin{array}{c}
x \, + \, i p_x \\
y \, + \, i p_y \end{array} \right ) \; \; {\rm with} ~
U \in SU(2)\times U(1). \;
\label{su2trans} \end{equation}

This symmetry has profound consequences on the spectrum of the system.
 The energy levels of this unperturbed system are trivially found and
  are given by
\begin{equation}
E_{N, m} \, = \, ( N \, + \, 1 ); ~ m \, =
\, N, N-2, N-4, \, \cdots \, , -(N-2) , -N \; ,
\label{hoeigen}\end{equation}
where $N$ is the principle quantum number (which is the sum of the
harmonic excitation quanta, $N = N_x + N_y$), and m is the
(two dimensional) angular momentum.  The interesting point is
that as a consequence of the symmetry, the levels are $(N+1)$-fold
degenerate.  In contrast, without the additional $SU(2)$ symmetry inherent
to the harmonic oscillator, the two-dimensional rotations imply
the $U(1)$ symmetry which requires that levels be two-fold
 degenerate for all $m \neq 0$.\\

Now suppose we add to the Hamiltonian a $SU(2)$ symmetry breaking
interaction (but which is still $U(1)$ invariant) of the form

\begin{equation}
V_{\rm SB} \, = \, A \, \theta (r - R),
\label{vsb}
 \end{equation}

\noindent
where  $A$ and $R$ are parameters and $\theta$ is the step
function.  Clearly, $V_{\rm SB}$ is not invariant under the
transformation of eq.~(\ref{su2trans}). Thus the $SU(2)$ symmetry
is explicitly broken by this additional interaction, that acts
only within a circle of radius $R$.
As a result one would expect that the eigenenergies will not
reflect the degeneracy structure of seen in eq.~(\ref{hoeigen}).
Of course, if either $A$ or $R$ is sufficiently small one expects
to find {\it nearly} degenerate multiplets (\ref{hoeigen})
everywhere in the spectrum.  However, if the
coefficients are sufficiently large one would expect to find no
obvious remnants of the multiplets low in the spectrum. Indeed,
 we have solved
numerically for the eigenstates for the case of $A=4$ and $R=1$
(in these dimensionless units) and one does not see an approximate
multiplet structure in the low lying spectrum as can be seen in
Fig.~1. This is not so surprising, the parameters were
chosen to be big enough to wipe out the ``would be'' degeneracy
structure.\\
\begin{figure}
\hspace*{-0.5cm}
$\begin{array}{cc}\psfig{file=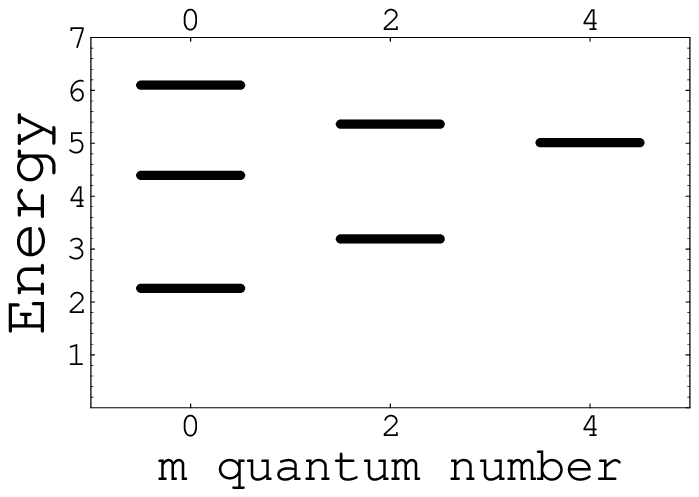,width=0.5\textwidth}
&\hspace*{-0.5cm}\psfig{file=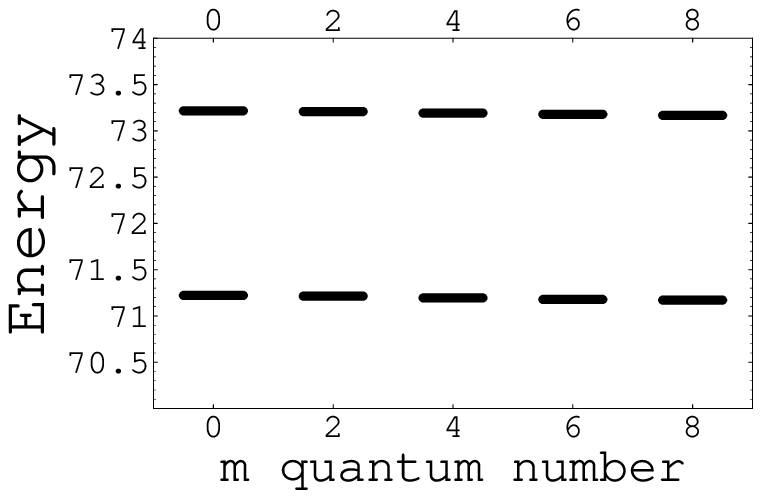,width=0.5\textwidth}\end{array}$
\caption{The low-lying (left panel) and highly-lying (right panel)
spectra of two-dimensional harmonic oscillator with the 
$SU(2)$-breaking term.}
\end{figure}

What is interesting for the present context is the high-lying spectrum.
  In Fig.~1 we have also plotted the energies between 70 and 74 for
   a few of the lower $m$'s (again for the case of $A=4$, $R=1$).
   A multiplet structure is quite evident---to very good approximation
    the states of different $m$'s form degenerate multiplets and,
    although we have not shown this in the figure these multiplets
     extend in $m$ up to $m=N$.  Thus the symmetry breaking
      interaction of eq~(\ref{vsb}) plays a dominant role in the
       spectroscopy for small energies and becomes insignificant
       at higher energies.  At higher energies, the spectroscopy
       reveals the $SU(2)$ symmetry of the two-dimensional harmonic
        oscillator.\\

How does this happen?  After all, the symmetry breaking term in
this case is explicitly there in the Hamiltonian so how is it
that it is not seen high in the spectrum?  In fact, it {\it is}
 seen high in the spectrum.  The highly-lying
 levels shown in Fig.~1 are
 {\it not} degenerate, they are merely almost degenerate.  The key
  point is that the effect of the symmetry breaking term is very small
   high in the spectrum and vanishes asymptotically high.
   Our central thesis is that something very
    similar happens for the case of spontaneous chiral symmetry
    breaking in QCD.\\

In the pedagogic case considered above, it is quite clear why
the symmetry
breaking term becomes unimportant high in the spectrum.
For very high lying states the energy of the state is much larger than the
 symmetry breaking interaction over the entire range
where the
symmetry breaking interaction acts and the wavefunction amplitude is very small
within the circle where the perturbation acts.  Thus, the wave function starting at the
origin and over entire space
 is, to very good approximation is simply
solution to the
 Schr\"odinger equation with the Hamiltonian  (\ref{HO}).
 The effect of the symmetry breaking
term then acts
as a perturbation

\begin{equation}
\Delta E_{N,m} = \langle N,m | V_{SB} | N,m \rangle.
\label{pert}
\end{equation}

\noindent
 Clearly
as $E \rightarrow \infty$ the effect of the symmetry breaking term on the
wave function and energies vanishes.\\ 

One thing about this simple problem is worth noting.  While the
effect of symmetry breaking becomes increasingly unimportant as
 one goes higher up in the spectrum, there is nothing discontinuous
 about the changes in the spectrum;  the spectrum smoothly changes
 from one regime to another.
 While there is a transition from the symmetry breaking
 regime low in the spectrum to the explicit symmetry regime
 asymptotically high in the spectrum, it should not be confused
 with a phase transition in the thermodynamic sense for a number of
reasons.  In the first place we are describing single states and not
an intensive quantity.  Secondly, the transition is smooth and to the
extent that it has a thermodynamic analog it would 
correspond to a cross-over
and not to a phase transition.  We expect that in QCD as in our toy
model there will be a gradual transition in the spectrum from the
low energy regime where spontaneous chiral-symmetry-breaking effects
are central to the physics to the high energy regime where they are
very small.\\

To summarize this section, we have shown an explicit example in two
dimensional
quantum mechanics where a system with an underlying symmetry is subjected to a
symmetry breaking effect (in this case an explicitly symmetry breaking
interaction) which destroys the multiplet structure in the spectrum associated
with the symmetry for low lying states but for which the high lying spectrum
retains the multiplet structure with nearly degenerate members of the
multiplet.
Our central argument is that something analogous happens in QCD:  a symmetry
breaking effect (in this case due to spontaneous symmetry breaking) affects
the
low-lying part of the spectrum and ruins the ``would be'' multiplet structure.
High in the spectrum however this symmetry breaking effect plays a very small
role and to good approximation the spectrum reflects the underlying chiral
symmetry.  As mentioned above, the analogy between the two cases is not
perfect---the QCD case depends on spontaneous symmetry breaking while the
simple
example is based on explicit symmetry breaking.  Moreover, in the QCD the
states of
 interest are in the continuum while in the toy problem they are discrete.
\\

  The distinction between
a continuous versus a discrete spectrum however raises critical issues.
Accordingly, in the following section we will discuss a tool, the spectral
density,
which enables us to ask sensible questions about a continuous spectrum.\\

\section{Spectral Densities, Correlation Functions and All That}

As noted above, the fact that all high lying states in the QCD spectrum
are in the continuum complicates any discussion about possible
 multiplets of hadrons high in the spectrum.  The difficulty is
 that there are states at any energy and thus some means must be
 found for differentiating between the states at different masses.
 To proceed, one must find some probe that samples these
 states and one can ask how strongly the probe couples to states at
 various masses.  This strength can be parameterized in a spectral
 density.  Before discussing in any mathematical detail how this is
 done a few general comments are in order.  The spectral density does
  not simply tell us about the states, it also tells about the probe,
  thus it may seem to be ill-suited to providing fundamental information
   about the spectrum.  However, there are cases where information
   about the spectrum more generally can be learned.  For example,
   if the system has discrete bound states then the spectral density
    represents a set of $\delta$ functions at  masses corresponding to the
    bound states.  While the strength multiplying the $\delta$ functions
    are properties of the probe as well as the states, the masses in
    the arguments of the $\delta$ function are independent of the
    probe and only tell you about the states in the spectrum.
    Unfortunately, the case of interest to us here is in the
    continuum so in principle one always contaminates
    the information about the underlying spectrum with the information
     about the probe.  However, if the spectrum is strongly resonant
      and the resonances are well separated
      then there will be a narrow region  with large spectral
      strength---{\it i.e.} a large bump.  Clearly the height of
      the bump will depend on the details of the probe however
      the position of the bump will be large insensitive.  There
      is some sensitivity of the position of the spectral bump to
      the probe as there is always some ambiguity in separating
      the resonance contribution from the background but for narrow
      resonances this ambiguity is small.  Thus, provided the spectrum
       of interest is in a region of well-defined resonances we can
       use any convenient probe and use the spectral density to
       determine the resonance position.\\

The easiest context to deal with a spectral density is the
two-point correlation function.   Consider a ``local current'',
$J(x)$ which we can construct entirely out of quark and gluon fields.
 We will choose $J(x)$ to be gauge invariant and to carry the
 quantum numbers we wish to study.   A useful listing of
 a number of such currents for various spin-flavor quantum
 numbers along with their transformation properties under
 chiral transformations is given in ref. \cite{CJ}. We
 can use $J$ to probe the vacuum---{\it i.e.} we act with
 $J$ on the vacuum and create the state with the quantum
 numbers of $J$,
 let this state propagate and act again with
 $J$ to bring the system back to the vacuum.  In the processes
 we learn about the propagation of all possible states with
 the quantum
  numbers of $J$.\\

Sometimes such a current is directly accessible experimentally.
For example, if one chooses the usual
electromagnetic current, one is able to study the
response of the vacuum to this vector probe, which is
represented by the excitations of vector mesons at low $s$ and
by jet production at the very high $s$. These vector
mesons which smoothly transform into jets once the
momentum transfer increases, are directly observable in the
process $e^+e^- \rightarrow
hadrons$. The total cross-section of this process is directly
connected to the imaginary part of the 2-point correlator
of the  current  in the time-like (Minkowsi) domain. The
experimental study of this process, in particular in the
regime $s \rightarrow \infty$, was historically one of the
most important arguments for  acceptance of QCD. Indeed, because
of asymptotic freedom, QCD predicts that in the regime
$s \rightarrow \infty$ the correlator is adequately described
by the free quark loop diagram (i.e. the photon creates from
the vacuum the quark-antiquark pair, which freely propagates
to the point where they are annihilated by the photon). The
spectral density is given by the imaginary part of this
diagram in the time-like region and is measurable as a total
cross-section. This process is described in standard texts
on QCD \cite{CL}.\\

If one uses the weak axial vector current, which is
experimentally accessible in decays of $\tau$-lepton,
then one studies the response of the vacuum to the axial-vector
probe; this is reflected in the excitations of axial
vector mesons, etc. For our purposes it is not necessary that the
spectral function associated with the current to
be experimentally accessible. For example,
there is no experimental probe of a local current that creates
three quarks from the vacuum. The two-point correlators with these currents give
information about the baryon spectrum. Such currents can be
constructed
theoretically \cite{IOFFE}, however, and they represent a tool to study
baryons in QCD sum rules and lattice gauge calculations \cite{SR,LATTICE}.\\

Consider for simplicity the two-point correlation function
for a Lorentz scalar (or pseudoscalar) $J$ defined as
\begin{equation}
\Pi_J(q^2) \, = \, i \, \int {\rm d}^4 x \, e^{- i q \cdot x} \langle 0
| \, T[ J(x) J(0)] |0 \rangle
\label{2pt}
\end{equation}
where $| 0 \rangle$ is the vacuum state and $T$ represents a
time-ordered product. This correlation function can be written
in standard K\"allen-Lehmann form \cite{KL}

\begin{equation}
\Pi_J(q^2) \, = \, - \int {\rm d}s \frac{\rho_J(s)}{q^2 - s + i \epsilon}
\label{kl}
\end{equation}
The spectral density $\rho_J(s)$ is defined as
\begin{equation}
\rho_J (s) \equiv \frac{1}{\pi} \rm{Im} \left ( \Pi_j(s) \right )
\label{rhoJ}
\end{equation}

\noindent
and has the physical interpretation of being proportional to the
probability density that the current $J$ when acting on the vacuum
creates a state of a mass of $\sqrt{s}$.  Analogous expressions for
nonscalars are slightly more complicated and will not be written
 down here but they can all be expressed using  the same general
 dispersive structure seen in eq.~(\ref{rhoJ}).\\

The key point from the present perspective is that the spectral
 density for two currents which are related to each other by chiral
  rotations become essentially equal high in the spectrum:

\begin{equation}
\lim_{s \rightarrow \infty} \left[ \rho_J(s) - \rho_{J'}(s)\right]
 \rightarrow 0
\label{speceq}
\end{equation}

\noindent
for $J' = U J U^{\dagger}$ where $U \in SU(2)_L \times SU(2)_R$ is 
a chiral rotation.\\

Equation (\ref{speceq}) is easily understood from considering
asymptotic properties of the correlation function at the large
{\it} space-like momenta and then using the dispersion relation
of eq.~(\ref{kl}) to relate this to the spectral density.  The
tool for calculating the correlator at large asymptotic $Q^2 =-q^2$
is the operator product expansion (OPE) \cite{OPE}. The operator product
$\int {\rm d}^4 x \, e^{- i q \cdot x} T[ J(x) J(0)]$ in this
regime can be written as the following operator series:
\begin{equation}
\int {\rm d}^4 x \, e^{- i q \cdot x} T[ J(x) J(0)] =
\sum_k C_k(Q^2, \alpha_s) {\cal O}_k,
\label{ope}
\end{equation}
where the Wilson coefficients, $ C_k(Q^2, \alpha_s)$ are calculable
in perturbation theory, the ${\cal O}_k$ are local gauge invariant
operators constructed from the quark and gluon fields and
$Q^2 = -q^2 >> \Lambda_{\rm QCD}$.  Examples of these ${\cal O}_J$
operators include identity operator
 $\1$, $\overline{q}q$ , $F_{\mu \nu}F^{\mu \nu}$ , {\it etc.}
 Thus the correlator can be expressed as
\begin{equation}
\Pi_J(Q^2) \, =  \, \sum_k C_k(Q^2, \alpha_s) \langle 0 | {\cal O}_k
| 0 \rangle,
\label{opecor}
\end{equation}
where the vacuum expectation values of the ${\cal O}_k$ are
referred to as ``condensates'' \cite{SVZ}.  In such an analysis all
 of nonperturbative effects including symmetry breaking
effects resides in the condensates.  The only effect that chiral
symmetry breaking can have on the correlator is through the nonzero
value of condensates associated with operators which are chirally
active ({\it i.e.} which transform nontrivially under chiral
transformations). To these belong $\langle  \bar q q \rangle$
and higher dimensional condensates that are not invariant under
axial transformation.\\

Simple dimensional analysis indicates that
\begin{equation}
\frac{C_m}{C_n} \sim Q^{{\rm dim}({\cal O}_n) - {\rm dim}({\cal O}_m)},
\end{equation}
so that at large  $Q^2$ the terms associated with high dimensional
operators are suppressed by  powers of $1/Q^2$.  This structure
in which the higher order condensates are increasingly suppressed by
 higher powers of $1/Q^2$ is essential to the usefulness of the OPE.
 At large $Q^2$ only a small number of condensates need be retained
 to get an accurate description of the correlator.  At asymptotically
 high $Q^2$, the correlator is well described by a single term---the
 perturbative term which multiplies the identity operator $\1$.  The
 essential thing to note from this OPE analysis is that the perturbative
 contribution knows nothing about chiral symmetry breaking as it contains
 no chirally nontrivial condensates. In other words, though the chiral
 symmetry is broken in the vacuum and all chiral noninvariant condensates
 are not zero, their influence on the correlator at asymptotically
 high $Q^2$ vanishes. This is in contrast to the situation of low values of
 $Q^2$, where the role of chiral condensates is crucial.
  Accordingly it is clear that if we
  consider two operators $J$ and $J'$ related to each other through chiral
   rotations then,
it must be true that for large $Q^2$ the two correlators become equal
 up to power law corrections.

\begin{equation}
\Pi_J(Q^2) - \Pi_{J'}(Q^2) \sim \frac{1}{Q^n} \; \; n>0 \; .
\label{highQ}
\end{equation}

The preceding relation merely shows that at large {\it space-like}
 momentum transfers the two correlation functions are identical,
 whereas the spectral functions describe the time-like region.
 However, the dispersion relation of eq.~(\ref{kl}) provides a
 connection between the space-like and time-like regions.  In
 essence one understands the fact that correlators for $J$ and
 $J'$ differ at low $Q^2$ and agree at large $Q^2$ up to power
 law corrections in the following way---the spectral densities
 agree (up to small corrections) at large $s$ and disagree only
 at small $s$. Such a structure guarantees the result of eq.~(\ref{highQ}).
 Thus one expects eqn.~(\ref{speceq}) to hold.  Strictly speaking
 at large $Q^2$ one does not require $\rho_J(s)$ to equal $\rho_{J'}(s)$
 on a point by point basis since at large space-like $Q^2$ one cannot
 resolve the fine structure at large $s$.  One does require however
 that coarse-grained integrals of the two must be essentially equal
 if integrated over moderate ranges in $s$.\\

Nevertheless, there are some limitations and ambiguities in the
procedure of analytical continuation from the deep Euclidean
domain to the Minkowski one \cite{SH}. Such a continuation were
unambiguous if only the function $\Pi_J(Q^2)$ had been known
exactly on some finite interval. In practice, however, one
always truncates the expansion (\ref{opecor}). In addition, there
could be also intrinsically nonperturbative contributions to
the coefficients functions $C_k$, e.g. a direct contribution
from the small size instantons (these nonperturbative contributions are
 however suppressed and do not contribute in the limit
$Q^2 \rightarrow \infty$).
 More importantly,
the OPE by itself does not determine the  function $\Pi_J(Q^2)$
everywhere. This is because the OPE is an expansion that picks
 only the light-cone singularities of the correlator. Thus
some possible singularities that are far from the light cone
intervals are not properly reflected in OPE. These singularities
could be quite important, however, in Minkowski domain. So
the question arises, what is a solid basis, nevertheless, for
arguing that the symmetries seen at large space like $Q^2$
will be reflected as symmetries in the spectral function.
In part, this reflects theoretical prejudice---it is hard to
imagine a situation in which it were to
fail grossly.  After all, as the
space-like $Q^2$ goes to $\infty$ one ``sees''
more and more of the  spectral function at large $s$ as seen from
the K\"allen-Lehmann
 representation of eq.~(\ref{kl}). Providing the correlator
does
not vanish in the $Q^2 \rightarrow \infty$ limit, the large $Q^2$
correlator will be totally dominated by the large $s$ spectral function
and one thus expects them to have the same symmetry behavior, justifying
eq.~(\ref{speceq}).  Moreover, there is an
  empirical basis for the validity of this type of argument. It is a well
established fact that, e.g. the deep inelastic processes (which
are sensitive to Euclidean kinematics) and the   process
$e^+e^- \rightarrow hadrons$ (that happens in Minkowski domain)
at $|q^2| \rightarrow \infty$ are both described by the same
free quark loop diagram which represents the first term in OPE \cite{GP}.
This free quark loop diagram is obviously insensitive to
spontaneous chiral symmetry breaking.\\

Thus, one sees that at large $s$ the spectral density for two
operators associated with each other via chiral rotations must
become the same.  Thus, the spectra at sufficiently large $s$ cannot
manifest any effects of chiral symmetry breaking and the spectral
densities for any chirally active currents must reflect a chiral
multiplet structure: the spectral strength for one channel
(corresponding to one current) must be very close to the spectral
strength for all channels related to via chiral transformations.
We can refer to the phenomenon of the spectral densities becoming
close as ``effective chiral restoration''.  We should note as we
 did in previous sections, this effective
 restoration is smooth and is not associated with a phase transition
 in the thermodynamic sense.  Rather it indicates that the effects of
 spontaneous chiral symmetry breaking on the spectrum are becoming
 progressively more irrelevant as one goes high in the spectrum so
 the symmetry is effectively restored.\\

 As noted in the beginning
of this section
the spectral density tells us about both the ``probe''
(in this case the current) and the spectrum.  It only gives
relatively unambiguous information about the spectrum independent
of the probe if the spectrum is strongly resonant.  Thus the
question we must address is whether the spectrum remains resonant
when one is high enough in the spectrum.  This subject will be
 discussed in the next section.\\

\section{Are There Hadronic Resonance Way Up There?}

The crux of the argument as outlined above is that from general 
considerations of the OPE in QCD, one deduces that the spectral 
densities at large $s$ for currents related by chiral transformations 
become identical.  This need not imply that hadrons form multiplets, 
however, since it may happen that by the time one is high enough in 
the spectrum for the spectral densities to be essentially equal, one 
is beyond the region of identifiable hadronic resonances.
Thus the conjecture that high in the spectrum of hadronic resonances 
there are multiplets associated with effective chiral restoration, comes 
down to the conjecture that the effects of chiral symmetry breaking 
on the spectrum turn off with increasing $s$ more or equally
rapidly than certain 
other nonpertubative effects---namely those effects responsible for 
the formation of hadronic resonances.\\

As a matter of principle it is therefore useful to demonstrate that it 
is at least possible that the effects of chiral symmetry breaking can die 
off in a region in which resonance are still well defined.  There is a 
very elegant argument due to Beane that demonstrates this.  The essence 
of this argument is that the phenomenon occurs for the problem of meson 
spectroscopy for QCD in the large $N_c$ limit.  As noted by `t Hooft in
his seminal paper \cite{Hooft},
 mesons become narrow in the large $N_c$ limit: three 
meson couplings scale as $N_c^{-1/2}$ implying mesonic widths which scale 
as $N_c^{-1}$ as $N_c \rightarrow \infty$, the widths go to zero and all 
mesons become stable. Thus, in the large $N_c$ limit, the dispersion integral 
of eq.~(\ref{kl}) becomes a discrete sum:
\begin{equation}
\Pi_J(q^2) \, = \, - \sum_l \frac{|a_l|^2}{q^2 - m_l^2 + i \epsilon}
\label{lm}
\end{equation}
where $l$ labels the meson and $a_l$ is the amplitude for the current 
acting on the vacuum to make the meson. It has long been known that the 
interplay between the perturbative results and the fact that mesons 
become narrow imposes strong constraints on the high part of the meson 
spectrum.\\ 

For example as noted by Witten \cite{Witten}, one immediately sees that as 
$N_c \rightarrow \infty$ there must be an infinite number of narrow mesons.  
The argument goes as follows. If there were only a finite number of terms 
in eq.~(\ref{lm}) then at large space-like $q^2$ the sum would fall off 
like $1/q^2$.  However, perturbation theory is valid in this region and 
perturbatively the correlation functions grow with increasing space-like 
$q^2$. These are incompatible and thus the hypothesis that only a finite 
number of mesons with finite mass contribute must be false.  Beane's argument is a variant 
of this: even if we are sufficiently high in the spectrum so that chiral 
symmetry breaking effects are unimportant, if $N_c$ is large enough 
eq.~(\ref{lm}) remains valid.  One key point in this is the fact the 
chiral symmetry breaking condensates do no grow with $N_c$ while 
the widths of the mesonic resonance decrease with $N_c$.  Thus, for the 
case of mesons in large $N_c$, the spectra indeed do form chiral multiplets 
of narrow resonant states if one goes sufficiently high in the spectrum
\cite{Beane}.  
This demonstrates by an explicit construction that it is {\it possible} for 
a system to be sufficiently
high in the spectrum so that 
chiral symmetry breaking effects become negligible 
while still being in a regime where the hadronic states are narrow.  
This large $N_c$ argument cannot be extended readily to the baryon 
spectrum since baryons do not become narrow in the large $N_c$ limit.\\

In fact, there is a small subtlety with the argument even in the mesonic case. 
The form of the spectral density in eq.~(\ref{lm}) is strictly valid for 
$N_c = \infty$.  For finite $N_c$ the mesons are narrow but, 
never-the-less are of finite width.  The widths of these states 
go as $N_c^{-1}$ since the coupling constants for three meson 
couplings go as $N_c^{-1/2}$ and the widths are proportional 
to the coupling constants squared.  However the proportionality 
constants depend on the available phase space for the decay and 
as one goes up in the spectrum the available phase for decay 
increases both since there are an increasing number of open decay 
channels and because each channel has larger phase space.  Thus 
the widths shrink with increasing $N_c$ but grow with increasing mass.  
Accordingly the behavior of the spectral density in the combined large $N_c$ 
and large $s$ limits depends on which limit one takes first.  For any 
mass and sufficiently large $N_c$ the mesons are narrow while for any 
$N_c$ for sufficiently large mass the mesons are wide.  This 
noncommutativity of limits might potentially be  an issue when one 
formulates a large $N_c$ argument along the line of the one given by Beane.  
Since the states of interest are high in the spectrum, how do we know that 
they are not so high as to be too wide to be isolated?  
The key point is that the chiral symmetry breaking effects do not 
grow with $N_c$ for large $N_c$ so we expect that high in the 
spectrum (where chiral symmetry breaking effects become insignificant) they 
are independent of $N_c$.  In contrast the value of $s$ where the mesons 
become so broad as to strongly overlap with other resonance of the same 
quantum numbers increases with increasing $N_c$.  
Thus one can always find an $N_c$ big enough so that 
the mesons are still narrow but chiral symmetry is effectively restored.\\

Beane's argument demonstrates that it is {\it possible} to be in a regime where 
hadrons are well-defined resonances while at the same time the effects of 
chiral symmetry breaking have become insignificant.  
The question, however, is does 
this happen in the real world of $N_c=3$?  Here we really do not have any 
reliable theoretical tools of calculations.  One might hope that 
eventually lattice 
QCD calculations may be able to shed light on this question. However all
 presently 
tractable formulations of lattice QCD are in Euclidean space and one must 
extrapolate this information into the time like region to learn about the 
spectrum. 
 This is straightforward for the lightest state with given quantum numbers
  but it 
is increasingly difficult as one goes up in the spectrum to separate out 
contributions of higher states.  Picking out resonances high in the spectrum 
is an 
intrinsically difficult task for lattice QCD. Thus for the foreseeable future
 we are 
unlikely to answer our question from {\it ab initio} calculations in QCD.  
Instead 
we will rely on an analysis of the experimental data
to see if there is evidence for this phenomenon.  This will be discussed 
in the 
following sections. \\

\section{Classification of the parity-chiral multiplets}

As discussed above, we know on very general grounds that
the effects of chiral symmetry breaking high in the
hadronic spectrum become small.  The key issue
which needs to be addressed is whether or not more-or-less
well-defined hadronic resonances exist in this regime.
Our conjecture is that for the baryon spectrum they do.
Now if this is  the case then high lying baryon
states will fall into multiplets of chiral group.  To
see if the experimental data supports such a conjecture
one must first determine what chiral multiplets are
expected and then see if the observed states fall into
these multiplets.\\

A simple way to generate the multiplets for baryon
states in the chirally restored regime is to use
a model in which the high-lying baryon is constructed
out of three  quark fields.
Such a scheme is simple, however it makes 
 model-dependent assumptions.  In particular, the construction
of a baryon out of three quark fields is common in 
constituent quark models.  However, the constituent quarks (which are massive -
 in contrast to current quarks - and do not belong
to any irreducible representation of the chiral group)
do not transform under the chiral group in the same manner
as current quarks.  Here, we assume three quark states {\it and}
quarks which transform chirally in the manner of massless current quarks.
For pedagogical purposes we will first
outline the classification scheme based on this model
 and only after that will describe a model-independent one.\\

Assume that baryon properties are determined by the properties
of the three valence quarks only. If chiral symmetry is
effectively  unbroken, then
the right and left components of valence quark fields are
decoupled

\begin{equation}
q=\frac{1-\gamma_5}{2}q + \frac{1+\gamma_5}{2}q \equiv q_l + q_r
\label{q}
\end{equation}

\noindent
and can be independently rotated in the flavor (isospin)
space. The irreducible representations of $SU(2)_L \times SU(2)_R$
  may be  labeled as $(I_L,I_R)$ where $I_L$ and $I_R$ represent the
isospin of the left- and right handed SU(2) groups. The left component
of the quark field is isodoublet with respect to the left
rotations, while it is not affected by right  rotations,
i.e. it is singlet (scalar) with respect to the right rotations.
Hence, the left component transforms as $(\frac{1}{2},0)$ irreducible
representation, while the right component transforms as $(0,\frac{1}{2})$
irreducible representation
of the chiral group. Consequently, according to (\ref{q}), the
one-quark field transforms as a direct sum of two irreducible representations

\begin{equation}
q \sim \left (\frac{1}{2},0 \right ) \oplus \left (0,\frac{1}{2} \right).
\label{qq}
\end{equation}

\noindent
We will refer such a representation as fundamental.\\

Since a baryon within this model picture consists of three
quarks only, the possible representations for baryon
in the chirally restored phase can
be obtained as a direct product of three fundamental representations
(\ref{qq}). Using the standard isospin coupling rules
separately for the left and right quark components, one
easily obtains a decomposition of this direct product

$$
\left[\left (\frac{1}{2},0 \right ) 
\oplus \left (0,\frac{1}{2} \right )\right]^3 =
\left[\left (\frac{3}{2},0 \right ) 
\oplus \left (0,\frac{3}{2} \right )\right]
$$

\begin{equation}
+ 3\left[ \left (1,\frac{1}{2}\right ) 
\oplus \left (\frac{1}{2},1\right )\right]
+ 3\left[\left (0,\frac{1}{2} \right ) 
\oplus \left (\frac{1}{2},0\right )\right]
+ 2\left[\left (\frac{1}{2},0 \right ) 
\oplus \left (0,\frac{1}{2} \right )\right]
.
\label{dec}
\end{equation}

\noindent
The last two representations in the expansion above
are identical group-theoretically, so they can be combined
with the common multiplicity factor 5.
Thus, according to the simple-minded model above,
baryons in the chirally
restored regime will belong to one of the
following representations
\begin{equation}
\left(\frac{1}{2},0 \right) \oplus \left(0,\frac{1}{2} \right);~
\left(\frac{3}{2},0 \right) \oplus \left(0,\frac{3}{2} \right);~
\left(\frac{1}{2},1 \right) \oplus \left(1,\frac{1}{2} \right).
\label{list}
\end{equation}

\noindent
As will be discussed below, the fact that parity is unbroken
by strong interactions restricts the baryon state
to a sum of two irreducible chiral representations.\\

The result (\ref{list}) is  essentially the correct one.
However, since it was obtained with strong model assumptions on the baryon
structure and since these assumptions are  questionable,
 it is important to develop a classification based
only on the fundamental symmetries and with no assumptions about
baryon structure \cite{TG}. This will be done in what follows.\\

Effective chiral symmetry restoration implies that the  physical
states must fill out representations of the chiral group
and transform into each other under chiral transformations.
At the same time all these physical states must be eigenstates of
parity, since the strong interaction does not break
parity.\footnote{In QCD there is a fundamental parameter $\Theta$,
and the parity is conserved if $\Theta=0$. The present phenomenological
data limits the possible value of this parameter  to at most to 
a very small
number. For purposes of strong interaction physics we can safely
set it to 0.}
Parity transforms  ``left'' into ``right'' and vice
versa. Hence, the states that fill out a general irreducible
representation $(I_a,I_b)$ cannot be at the same time
eigenstates of the parity, because under parity transformation
those states transform into the states that belong to a
different irreducible representation,

\begin{equation}
P |(I_a,I_b)\rangle = |(I_b,I_a)\rangle.
\label{p}
\end{equation}

\noindent
Irreducible chiral representations are invariant
under parity transformations only
for the case $I_a=I_b$.  However,
the states in the representation $(I_a,I_a)$ only have
integral isospin in the range $I =0,1,...,2I_a$ and thus
cannot be baryons in two flavor QCD. (Recall that with
two flavors baryons must have
a half integral isospin). Thus multiplets must
correspond to reducible representation of the chiral group.
We have to construct the minimal possible representations of the chiral group
for half-integral isospin that are compatible with
definite parity for the states. This task is simple
because there is an automorphism of the group $SU(2)_L \times SU(2)_R$
with respect to a mapping of the left and right subgroups.
This mapping is an interchange of the left and right
chiral charges, $Q^i_L \leftrightarrow Q^i_R$, where $i$
refers to isospin projection. Under this operation the
vector charge (isospin) $Q^i =Q^i_L + Q^i_R$ is not affected,
while the axial charge, $Q^i_5 = Q^i_R - Q^i_L$, changes its sign.
Since such an operation is the parity transformation, the minimal
possible representation of the chiral group that is invariant
under parity operation (i.e. under parity transformation
every state in the given representation transforms into a
state within the same representation) must contain
two distinct irreducible representations of $SU(2)_L \times SU(2)_R$
that transform into each other under parity operation

\begin{equation}
 (I_a,I_b) \oplus (I_b,I_a).
\label{r}
\end{equation}

\noindent
We refer such a representation as a {\it parity-chiral}
multiplet. When chiral symmetry is (almost) restored in spectral density
and if the identifiable baryon resonances still exist, then the baryons high in the
spectrum fall into such multiplets.
 While this representation is a {\it reducible} one
with respect to the chiral group $SU(2)_L \times SU(2)_R$,
it is an {\it irreducible} one with respect to the wider symmetry group

\begin{equation}
 SU(2)_L \times SU(2)_R \times C_i,
\label{gr}
\end{equation}

\noindent
where the group $C_i$ consists of two elements: identity
and inversion in 3-dim space.\footnote{
In the literature language is sometimes used in
a sloppy way and the representation (\ref{r})
is referred to erroneously as an irreducible representation of the
chiral group.}
This symmetry group is
the symmetry of the QCD Lagrangian (neglecting quark masses),
 however only its
subgroup $SU(2)_I \times C_i$ survives in the broken symmetry mode.
The dimension of the representation (\ref{r}) is

\begin{equation}
dim_{ (I_a,I_b) \oplus (I_b,I_a)} = 2(2I_a+1)(2I_b+1).
\label{dim}
\end{equation}

The isospin group $SU(2)_I$ is a subgroup of the chiral group,
and the isospin symmetry survives in the broken chiral symmetry
mode. Hence the isospin is a good quantum number that can be used to
classify states in both (approximately) restored and broken chiral
symmetry regimes.
In the explicit chiral symmetry regime the isospin of the
state can be obtained from the left and right isospins according
to a standard angular momentum rules. The given representation
of the parity-chiral group contains states with {\it all} possible
isospins

\begin{equation}
I=|I_b-I_a|, |I_b-I_a| +1,..., I_b+I_a.
\label{i}
\end{equation}

\noindent
The states of definite parity in the chirally restored regime
are constructed from the states with definite chirality and
isospin, that we denote $|I_{(I_a,I_b)}\rangle$.
  The states of positive and negative parity are

\begin{equation}
2^{-1/2} \left ( |I_{(I_a,I_b)}\rangle +  P |I_{(I_a,I_b)}\rangle \right )
= 2^{-1/2} \left ( |I_{(I_a,I_b)}\rangle +  |I_{(I_b,I_a)}\rangle \right )
\label{pos}
\end{equation}

\noindent
and

\begin{equation}
2^{-1/2} \left ( |I_{(I_a,I_b)}\rangle -  P |I_{(I_a,I_b)}\rangle \right )
=2^{-1/2} \left ( |I_{(I_a,I_b)}\rangle -   |I_{(I_b,I_a)}\rangle \right ),
\label{pos}
\end{equation}

\noindent
 respectively.\\

Empirically, there are no known baryon resonances within the two
light flavor sector which have an isospin greater than 3/2.
Thus we have a constraint from the data that
if chiral symmetry is effectively restored for very highly
excited baryons, the only possible representations for the
 observed baryons have $I_a + I_b \le 3/2$, {\it i.e.} the
 only possible representations are
$(1/2,0) \oplus (0,1/2)$, $(1/2,1) \oplus (1,1/2)$
 and $(3/2,0) \oplus (0,3/2)$. Since chiral symmetry and
parity do not constrain the possible spins of the states
these multiplets can correspond to states of any fixed spin.
Note this empirical constraint reduces the allowable representations
to precisely those seen in the simple valence quark model discussed
above.
It is worth noting that the constraint on the allowable representations
stemming from the lack of known resonances with isospin larger than
3/2 is somewhat weak.  It is conceivable that there do exist baryon
resonances with $I > 3/2$ that simply have not yet been observed.
If such states do exist, they greatly expand the possible multiplets.
In the analysis that follows we assume that these states either do
not exist as well-defined resonances or do so at an energy above
the present data (where picking any resonance out of the data
is hard).  This constrains the allowable multiplets to the
ones enumerated above.  However, if there were an unidentified
resonance with $I>3/2$ with an energy comparable to the
known resonances, it is possible that our assignments of
states to multiplets would need to be altered.\\

The $(1/2,0) \oplus (0,1/2)$ multiplets contain only isospin
1/2 states and hence correspond to parity doublets of nucleon
states (of any fixed spin).\footnote{If one distinguishes
nucleon states with different electric charge, i.e. different
isospin projection, then this ``doublet'' is actually a quartet.}
Similarly, $(3/2,0) \oplus (0,3/2)$
 multiplets contain only isospin 3/2 states and hence correspond
to parity doublets of $\Delta$ states (of any fixed spin).\footnote{
Again, keeping in mind different charge states of delta
resonance it is actually an octet.}
However, $(1/2,1) \oplus (1,1/2)$ multiplets contain both
isospin 1/2 and isospin 3/2 states and hence correspond to
 multiplets containing both nucleon and $\Delta$ states of
both parities and any fixed spin.\footnote{This representation
is a 12-plet once we distinguish between different charge states.}\\

Summarizing, the phenomenological consequence of the effective
restoration of chiral symmetry
high in $N$ and $\Delta$ spectra is that the baryon states
will fill out  the irreducible
representations of the parity-chiral group (\ref{gr}).
If $(1/2,0) \oplus (0,1/2)$ and $(3/2,0) \oplus (0,3/2)$
multiplets were realized in nature, then the spectra of highly excited
nucleons and deltas would consist of parity doublets. However,
the energy of the parity doublet with  given spin in
the nucleon spectrum {\it a-priori} would not be degenerate with the
the doublet with the same spin in the delta spectrum;
these doublets would belong to different
representations of eq.~(\ref{gr}), {\it i.e.} to distinct
multiplets and their energies
are not related.   On the other hand,
if $(1/2,1) \oplus (1,1/2)$ were realized, then the highly
lying states in $N$ and $\Delta$ spectrum 
would have a $N$ parity doublet and a $\Delta$
parity doublet with the same spin and which are degenerate in mass.
In either of cases the highly lying spectrum  must systematically
consist of parity doublets.\\

We stress that this classification is the most general one
and does not rely on any model assumption about the
structure of baryons. The only assumption beyond those of
effective symmetry restoration and the lack of parity breaking
is that the states fall into representations with $I \le 3/2$.
This last constraint is empirical in nature.

\section{Review of the experimental data}

Before reviewing the experimental situation in detail, a
few words of caution should be given.  We rely on the Particle
Data Group's \cite{PDG} compilation of the resonances and we use the masses
of these resonances in attempting to assess whether states are
nearly degenerate.  It is worth noting at the outset however, that strictly
speaking, the resonance masses reported by the PDG are not experimental
quantities.  The actual experimental quantities are various scattering
observables such as differential cross-sections.  The resonance parameters
can only be extracted from these observables via some type of modeling.
For example an amplitude written as the sum of a resonant contribution
plus a background term of some prescribed form.  Clearly, there
is some model dependence in the extraction of the parameters so technically
the extracted masses are not purely experimental quantities.  However,
for strong resonances the model dependence is weak.\\

\begin{figure}
\psfig{file=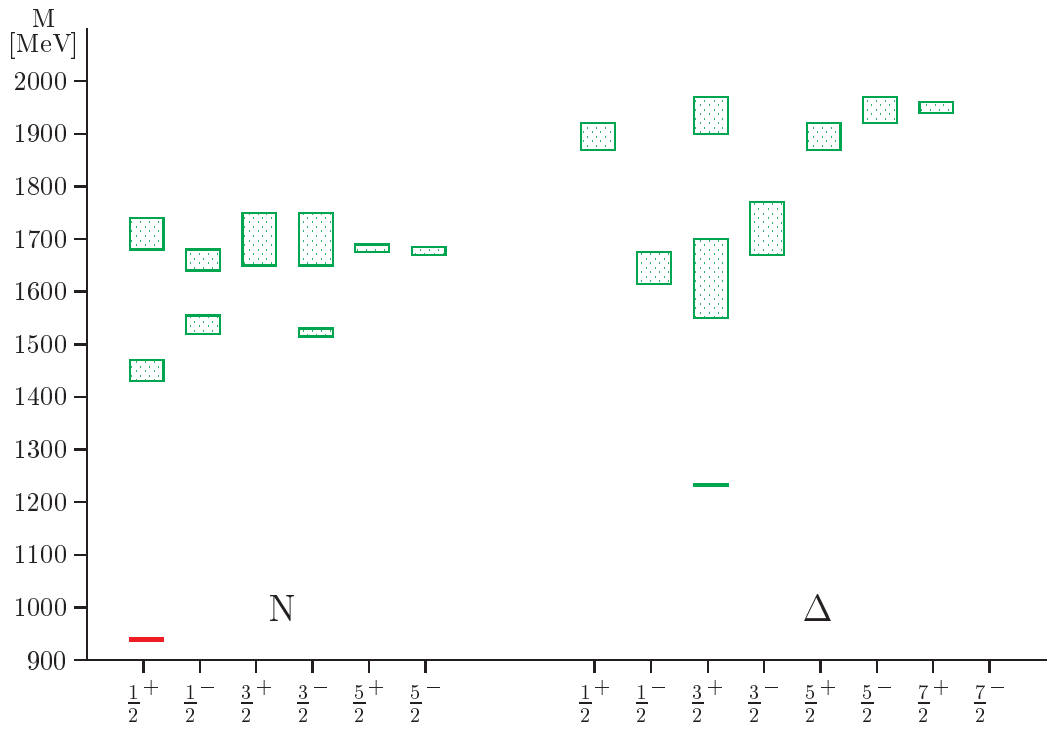}
\caption{The low-lying  $N$ and $\Delta$ experimental spectra.
The shadowed boxes represent experimental uncertainties for
baryon masses.}
\end{figure}

In Fig. 2 we show all the well established states in
$N$ and $\Delta$ spectra below 2 GeV.  Up to approximately
1.8 GeV the spectrum is well explored experimentally.
However, this is not the case higher in the spectrum and
it would not be surprising if future experiments or
reanalysis of old data yield
 the new resonant states.\\

What is immediately evident from the low-lying spectrum is 
that  positive and negative
parity states with the same spin are not nearly degenerate.
Even more, there is no one-to-one mapping of positive and
negative parity states of the same spin with masses below 1.8 GeV.
This means that one cannot
describe the low-lying spectrum as consisting of sets of chiral 
partners.\footnote{There are attempts in the literature to classify
the low-lying baryon resonances into representations of the
chiral group in the spirit of the $\sigma$-model. In order these
attempts be successful it is necessary to have a one-to-one
mapping of baryon states in the Wigner-Weyl mode (where the whole
spectrum would represent a set of degenerate parity-chiral
multiplets) into the states in the actual world (i.e. into the
physical states in the Nambu-Goldstone mode), which is 
not observed. There is a fundamental reason why such attempts
cannot be grounded in QCD where hadrons are composite objects. 
It could be correct if there were
a continuous smooth transition for {\it all} hadron states from the
Nambu-Goldstone mode to the Wigner-Weyl one. This is in
conflict, however, with the Coleman-Witten theorem \cite{CW} which states
that in the large $N_c$ limit in the {\it confining} regime 
({\it i.e.} in the regime where all hadrons as color-singlet
entities still exist)
the
chiral symmetry must be spontaneously broken to the vectorial
subgroup. Hence at least in the large $N_c$ limit in QCD it is not
possible to provide at the same time the existence of all hadrons 
and the explicit (Wigner-Weyl) mode of chiral symmetry. This
then implies that at least in the large $N_c$ limit in QCD
there is no continuous smooth transition for all hadron states
from the spontaneously broken to the explicit mode of chiral
symmetry. In other words, it is impossible to define the
baryon fields that are classified into chiral multiplets
and that one-to-one map into physical baryons.
 This is also consistent with the standard wisdom that
transition from the Nambu-Goldstone mode to the Wigner-Weyl
one is a phase transition that is a-priory
discontinuous.}\\

The absence of  systematic parity doublets low in the spectrum
 is one of the most direct pieces of evidence that chiral symmetry in QCD is
spontaneously broken. However, as follows from the discussion in
previous sections, there are good reasons to expect that chiral
symmetry breaking effect become progressively less important
higher in the spectrum. As a phenomenological manifestation
of this smooth chiral symmetry restoration one should expect
an appearance of  systematic parity-chiral multiplets high in
the spectrum.\\

The question of relevance is whether the observed baryon
 highly lying resonances fall into these representations.
This is not
 trivial to determine for a number of reasons.  The first
 is that the  theoretical arguments
discussed in the previous section suggest that the effective
 chiral restoration is only approximate  both because of finite
quark mass
effects and due to residual effects of spontaneous symmetry breaking
which we have argued are expected to turn off gradually.
  Moreover, we have no tools to estimate in an {\it a priori}
 fashion the expected size of these symmetry-breaking effects
 high in the baryon spectrum.  Thus some judgment is need to
 assert that two levels are ``nearly degenerate''.  A second
 complication stems from the fact that this high in the spectrum
 there are many levels close together and one cannot
rule out the possibility that two states are near each other
in energy by accident.  Moreover, as noted above the
extraction of  the ``experimental data''  in \cite{PDG} is somewhat
model dependent. In addition high in the spectrum some  resonant
states may well have been missed
and some states may have masses which have been misestimated.\\

Keeping all these in mind, we note however, that the
known empirical spectra of the highly lying $N$ and
$\Delta$ baryon resonances  suggest remarkable regularity.
Below we show all the known $N$ and $\Delta$ resonances in the
region 2 GeV and higher and include not only the well
established baryons (``****'' and ``***'' states according to the
PDG classification \cite{PDG}), but also ``**'' states that are defined
by PDG as states where ``evidence of existence is only fair''.
In some cases we will fill in the vacancies in the classification
below by the ``*'' states, that are defined  as
``evidence of existence is poor''. We  mark both the 1-star
and 2-star states in the classification below.

$${\bf J=\frac{1}{2}:}
 ~N^+(2100)~(*),~N^-(2090)~(*),~\Delta^+(1910)~~~~,~\Delta^-(1900) (**);$$

$$ {\bf J=\frac{3}{2}:}
 ~N^+(1900) (**),~N^-(2080) (**),~\Delta^+(1920)~~~~,~\Delta^-(1940)~(*);$$

$${\bf J=\frac{5}{2}:}
 ~N^+(2000) (**),~~N^-(2200) (**),~ \Delta^+(1905)~~~~,~\Delta^-(1930)~~~~;$$

$${\bf J=\frac{7}{2}:}
 ~N^+(1990) (**),~ N^-(2190)~~~~,~\Delta^+(1950)~~~~,~\Delta^-(2200)~(*);$$

$${\bf J=\frac{9}{2}:}
 ~N^+(2220)~~~,~N^-(2250)~~~~,~\Delta^+(2300) (**),~\Delta^-(2400)(**);$$

$${\bf J=\frac{11}{2}:}
 ~~~~~~~~~?~~~~~~~~,~N^-(2600)~~~~,~\Delta^+(2420)~~~~,~~~~~~~~?~~~~~~~~;$$

$${\bf J=\frac{13}{2}:}
 ~N^+(2700) (**), ~~~~~~~?~~~~~~~~,~~~~~~~~?~~~~~~~~,~\Delta^-(2750)(**);$$

$${\bf J=\frac{15}{2}:}
 ~~~~~~~~?~~~~~~~~,  ~~~~~~~~?~~~~~~~~,~\Delta^+(2950) (**),~~~~~~~~?~~~~~~~~.$$

The data above suggest that the parity doublets in $N$
and $\Delta$ spectra are approximately degenerate; the typical splitting
in the multiplets are $\sim 200$ MeV or less, which is within
the decay width of those states.  Of course, as noted
above,``nearly degenerate'' is not a truly well-defined idea.
In judging how close to degenerate these states really are
one should keep in mind that the extracted resonance masses have
uncertainties which are typically of the order of 100 MeV.\\

We stress that the 1-star-states by no means should be
taken very seriously. The uncertainty interval for the
masses of 1-star and 2-star should be taken essentially
bigger, than for the well established states. This is
because the masses of these states were extracted from
rather old phase shift analysis and the results of
different groups often do not coincide with each other.\\

Though one cannot  rule out the possibility that the approximate
mass degeneracy between the $N$ and $\Delta$ doublets is
accidental ( which would presumably mean  that that the baryons are organized
according to $(1/2,0) \oplus (0,1/2)$ for $N$ and
$(3/2,0) \oplus (0,3/2)$ for $\Delta$ parity-chiral doublets),
we believe that
this fact supports the idea (ii) that the highly excited states
fall into approximately degenerate multiplets
$(1/2,1) \oplus (1,1/2)$.\\

It can also be possible that in the narrow energy interval
more than one parity doublet in the nucleon and delta spectra
is found for a given spin. This would then mean that different
doublets would belong to different parity-chiral multiplets.\\

While a discovery of states that are marked by (?)
would support the idea of effective chiral symmetry restoration,
 a definitive discovery of states that are beyond the systematics of
parity doubling, would certainly be strong evidence against it.
The nucleon states listed above exhaust all states
(``****'', ``***'', ``**'', ``*'') in this part of the spectrum
included by the PDG. However, there are some additional
candidates (not established states) in the $\Delta$
spectrum. In the $J=5/2$ channel
there are two other candidate states $\Delta^+(2000)(**)$ and
$\Delta^-(2350)(*)$; there is another
candidate for $J=7/2$ positive parity state - $\Delta^+(2390)(*)$
as well as  for $J=1/2$ negative parity
state $\Delta^-(2150)(*)$. Certainly a better exploration
of the highly lying baryons is needed. This task is just for
the facilities like in JLAB, BNL, SAPHIR, SPRING-8 and similar.\\

Recent experimental data from SAPHIR (Bonn) \cite{BONN} indicate
two additional states in the nucleon spectrum:

$${\bf J=\frac{1}{2}:}
~N^+(1986 \pm 26^{+10}_{-30}),N^-(1897 \pm 50^{+30}_{-2}).$$

\noindent
What is interesting, the states again appear as approximate
parity doublets. It is not clear at the moment whether or not
these new states should be actually identified with those states
above mentioned in PDG.\\

\section{Conclusion and outlook}

 In this short review we have shown that the
 chiral symmetry of QCD must be 
restored smoothly as one goes up in the hadron spectra.
 The arguments are based  very general
 properties such as  quark-hadron duality,
  asymptotic freedom in QCD and the validity
 of the operator product expansion in QCD. Using
 these we have demonstrated that asymptotically high
 in the spectrum, the spectral densities obtained
 with  local currents that transform into each
 other under chiral transformations, must coincide.
 This is in  marked contrast to the low-lying part
 of the spectra where these spectral densities are
 very different due to the spontaneous breaking
 of chiral symmetry of the vacuum. Physically this
 is quite easy to understand. The chiral symmetry breaking
 of the vacuum is simply not important 
 high in the spectrum, while it is crucial for the
 low-lying states.\\
 
  To the extent
 that the identifiable hadronic resonances still
 exist in the continuum spectrum at high excitations
 the effective chiral symmetry restoration in the spectral
 densities implies that the highly excited hadrons
 should fall into multiplets associated with
  the representations of the chiral
 group. In other words, the spectrum of highly excited
 $N$ and $\Delta$ states should consist 
 of the approximate parity doublets or higher multiplets.
 There are two possibilities: (i) the parity doublet in the
 nucleon spectrum is not degenerate with the doublet of
 the same spin in the delta spectrum. This would imply
 that these doublets belong to different chiral multiplets --
 to $(1/2,0) \oplus (0,1/2)$ for $N$ and to
$(3/2,0) \oplus (0,3/2)$ for $\Delta$;
 (ii) the parity doublets of the same spin in nucleon and
 delta spectra are degenerate. Then it would support the
 possibility that both parity doublets belong to the same
 multiplet -- $(1/2,1) \oplus (1,1/2)$. 
  The small splitting
 in these doublets is due to the explicit chiral symmetry
 breaking by small masses of $u$ and $d$ quarks as well as
 due to the remaining small effects of spontaneous symmetry
 breaking. The latter effects vanish completely only
 asymptotically high, where the identifiable hadrons
 probably do not exist in the real world.\\

 We have shown that the existing old data on highly
 excited $N$ and $\Delta$ baryons in the region of
 2 GeV and higher do support this picture. However
  new experimental studies are necessary to reach
  definitive conclusion on whether 
  nature realizes approximate
 chiral symmetry restoration in this region. Such studies
 could  be performed with the existing facilities.\\

 One question which arises is why do we see the chiral symmetry
 restoration in baryon spectrum  but
  no such evidences in meson
 spectrum? Indeed, the approximate chiral symmetry
 restoration would require that e.g. the highly excited
 vector and axial vector mesons also form  approximate
 parity doublets. If one looks at the PDG tables, then
 one finds the vector mesons at the following masses
 $\rho(770)$, $\rho(1450)$, $\rho(1700)$, $\rho(2150)$.
 There are, however, only two axial vector meson states: $a_1(1260)$
 and $a_1(1640)$. 
 As expected from spontaneous chiral symmetry breaking,
 there is no parity doubling low in the spectrum.
  (Compare the mass of $\rho(770)$ and $a_1(1260)$).
 However, there is no hint of parity doubling high in the spectrum
 at masses of 2 GeV, as one sees in the baryon spectrum. 
 One possible reason
 is trivial however -- with the present experimental possibilities
available to date
 one might expect to have trouble seeing such a doubling
 in meson spectrum.
 In the case of baryons, the nucleon targets exist and
 one can perform direct experiments on high excitation of
 nucleons by different projectiles such as pions, protons, electrons or
 photons. The multipole analysis needed to extract resonances
 in various channels is relatively straightforward. On the other
hand,   the meson targets do not exist;
except in a few cases 
 the only way to extract the meson spectrum is from
 indirect experiments. For example, the vector mesons are
 obtained directly from the $e^+e^-$ annihilation process, which is
 well explored and the $\rho$ mesons are indeed known up to
 2 GeV region. The extraction of higher vector mesons is
 more difficult due to the opening of the hidden
 charm production channel. The axial vector mesons are
 obtained directly from the weak decay of the $\tau$-lepton. The
 mass of $\tau$ is 1777 MeV, which guarantees that we could
 not observe the possible axial vector mesons in the
 2 GeV region. Other types of experiments are necessary.\\


\begin{thebibliography}{99}
\bibitem{G1} L. Ya. Glozman, Phys. Lett. {\bf B475},329 (2000).
\bibitem{TG} T.D. Cohen and L. Ya. Glozman, hep-ph/0102206;
 Phys. Rev. {\bf D65}, 016006 (2002).
\bibitem{GML} M. Gell-Mann and M. Levy, Nuovo Cimento {\bf 16} 705 (1960).
\bibitem{ADDA} S. L. Adler and R. F. Dashen, {\it Current Algebras
And Applications to Particle Physics}, W. A. Benjamin, Inc;
New York, 1968.
\bibitem{P} H. Pagels, Phys. Rep. {\bf 16}, 219 (1975).
\bibitem{Collins} P. D. B. Collins, {\it An Introduction to Regge
Theory and High Energy Physics}, Cambridge University Press, Cambridge,
1977.
\bibitem{GREGGE} L. Ya. Glozman, hep-ph/0105225.
\bibitem{GR} L. Ya. Glozman and D.O. Riska, Phys. Rep., {\bf 268}, 263 (1996).
\bibitem{Beane} S. R. Beane, hep-ph/0106022, Phys. Rev. {\bf D64},
 116010 (2001).
\bibitem{VW} C. Vafa and E. Witten, Nucl. Phys. {\bf B234}, 173 (1984).
\bibitem{GOR} M. Gell-Mann, R. J. Oakes and B. Renner,
 Phys. Rev. {\bf 175}, 2195 (1968).
\bibitem{CL} T.P. Cheng and L.F. Li, {\it Gauge Theory of
Elementary Particle Physics}, Clarendon Press, Oxford, 1984.
\bibitem{IOFFE} B. L. Ioffe, Nucl. Phys. {\bf B188}, 317 (1981);
E:{\bf B191}, 591.
\bibitem{SR} P. Colangelo and A. Khodjamirian, "QCD Sum Rules,
a Modern Perspective", in: At the Frontier of Particle Physics/
Handbook of QCD, ed. M. Shifman, World Scientific, Singapore, 2001.
\bibitem{LATTICE} I. Montvy and G. M\"unster, ``Quantum Fields
on the Lattice'', Cambridge University Press, Cambridge, 1994.
\bibitem{OPE} K. G. Wilson, Phys. Rev. {\bf 179}, 1499 (1969).
\bibitem{SVZ} M. A. Shifman, A. I. Vainstein, and V. I. Zakharov,
Nucl. Phys. {\bf B147}, 385 (1979).
\bibitem{GP} H. Georgi and H. D. Politzer, Phys. Rev. {\bf D9}, 416 (1974).
\bibitem{IOFFE} B. L. Ioffe, Nucl. Phys. {\bf B188}, 317 (1981);
E:{\bf B191}, 591.
\bibitem{SH} M. Shifman, "Quark-Hadron Duality", in:
At the Frontier of Particle Physics/
Handbook of QCD, ed. M. Shifman, World Scientific, Singapore, 2001.
\bibitem{CJ} T. D. Cohen and X. Ji, Phys. Rev. {\bf D55}, 6870 (1997).
\bibitem{KL} G. K\"allen, Hev. Phys. Acta {\bf 25},417 (1952); H. Lehmann,
Nuovo Cumento {\bf 11} 342 (1954).  This topic is discussed in standard field
theory texts.  See for example S. Weinberg, {\it The Quantum Theory of
Fields}, Cambridge University Press, Cambridge, 1995.
\bibitem{Hooft}G. 't Hofft, Nucl. Phys. {\bf B72}, 461 (1974).
\bibitem{Witten}  E. Witten, Nucl. Phys. {\bf B156}, 269 (1979).
\bibitem{CW} S. Coleman and E. Witten, Phys. Rev. Lett. 
{\bf 45}, 100 (1980).
\bibitem{PDG} Particle Data Group, Europ. Phys. Journ. {\bf C15},1 (2000).
\bibitem{BONN} R. Pl\"otzke et al, Phys. Lett. {\bf B444} 555 (1998).
\end{thebibliography}
\end{document}